\documentclass[%
 ,twocolumn%
,unsortedaddress%
,aps, prd
]{revtex4}
\usepackage{bm}%

\begin{document}

\newcommand{\abs}[1]{\left| #1 \right|}
\newcommand{\nn}[0]{\nonumber \\ }

\preprint{\vbox{\hbox{UCSD/PTH 02-15}}}

\title{$\Delta \to N \gamma$ in Large-$N_c$ QCD}%

\author{Elizabeth Jenkins}%
\email{ejenkins@ucsd.edu}
\affiliation{University of California, San Diego\\
9500 Gilman Drive, La Jolla, CA 92093-0319}

\author{Xiangdong Ji}%
\email{xji@physics.umd.edu}
\affiliation{Department of Physics, University of Maryland, 
\\ College Park, MD 20742}

\author{Aneesh V. Manohar}%
\email{amanohar@ucsd.edu}
\affiliation{University of California, San Diego\\
9500 Gilman Drive, La Jolla, CA 92093-0319}

\begin{abstract}
The decay $\Delta^+ \rightarrow p \gamma$ is studied in the $1/N_c$ expansion
of QCD. The ratio of the helicity amplitudes is determined to be
$A_{3/2}/A_{1/2} = \sqrt{3}  + {\cal O}(1/N_c^2)$.  Equivalently,    the ratio
$E2/M1$ of the multipole amplitudes is predicted to be order $1/N_c^2$.
\end{abstract}
\date{July 2002}%
\maketitle

One of the mysteries in hadronic physics is the success of the
non-relativistic quark model in describing the structure and spectrum of the
low-lying baryons. In the non-relativistic quark model, the spin structure of
baryons is determined entirely by the spin of the constituent quarks, in clear
contradiction to the result from polarized deep-inelastic scattering that only
20-30\% of the nucleon spin is carried by the spin of the quarks 
\cite{filippone}. Nevertheless, the non-relativistic quark model 
successfully predicts that the ratio of the neutron to proton magnetic moments is $-2/3$  
and that the $\Delta$ to $N$ electromagnetic transition is dominated by the M1 multipole.  
Recent advances in studying baryons in the $1/N_c$ expansion of QCD help to
reconcile the success of the quark model predictions with the spin structure
of the nucleon~\cite{rev1,rev2,dmj,djm}.

The large-$N_c$ limit of QCD validates many of the quark model spin-flavor
symmetry results without assuming a non-relativistic quark dynamics for the
baryon wave functions.  An  SU(6) spin-flavor symmetry of baryons was derived
in the large-$N_c$ limit, which yields the same spin-flavor results as the
non-relativistic quark model and Skyrme model without making any of  the
dynamical assumptions of these models. In the $1/N_c$ expansion of QCD,  the
ratio of the neutron and proton magnetic moments is predicted  to be $-2/3$ up
to a correction of order $1/N_c^2$~\cite{rev1,rev2,dmj,djm}. Likewise, it is
shown in this paper that the E2/M1 electromagnetic transition ratio between
$\Delta$ and $N$ is of order $1/N_c^2$.  This result requires no assumption
about the underlying quark orbital angular momentum, or intrinsic deformation,
of the baryon.

The radiative decay $\Delta^+ \to p \gamma$ is described by two independent 
amplitudes. The standard convention~\cite{PDG}  is to use helicity amplitudes
$A_{1/2}$ and $A_{3/2}$, which are  the amplitudes for a $\Delta^+$ state with
helicity $1/2$ or $3/2$ along its direction of motion  to decay to a
right-handed photon moving along this same direction, and a proton  recoiling
in the opposite direction. The helicity amplitudes for $\Delta^+$ decay to a
left-handed photon are related to $A_{1/2}$  and $A_{3/2}$ by parity
conservation of the strong and electromagnetic  interactions.  The total decay
rate  for $\Delta^+ \to p \gamma$ is given by
\begin{eqnarray}
\Gamma &=& {k_\gamma^2 \over \pi} { 2 M_p \over (2J+1) M_\Delta} 
\left[ \abs{A_{1/2}}^2 +
\abs{A_{3/2}}^2 \right] \ ,
\label{100}
\end{eqnarray}
where $k_\gamma=259$~MeV is the momentum  of the photon, $M_p$ is the proton 
mass, $M_\Delta$ is the $\Delta^+$ mass, and $J=3/2$ is the spin of  the
$\Delta^+$.  Eq.~(\ref{100}) adheres to the normalization convention of the
helicity amplitudes used by the Particle Data Group~\cite{PDG}.  The
experimental values for the helicity amplitudes are
\begin{eqnarray}
A_{1/2} &=& -0.135 \pm 0.006 \ \text{GeV}^{-1/2}, \nn
A_{3/2} &=& -0.255 \pm 0.008\ \text{GeV}^{-1/2},
\label{200}
\end{eqnarray}
which yield the ratio
\begin{eqnarray}
{A_{3/2}  \over A_{1/2} } &=& 1.89 \pm 0.10 \ .
\label{300}
\end{eqnarray}
In this paper, we show that the large-$N_c$ prediction for this ratio is
\begin{eqnarray}
{A_{3/2}  \over A_{1/2} } &=& \sqrt{3} + \mathcal{O}
\left( { 1 \over N_c^2} \right) =1.73 + \mathcal{O}
\left( { 1 \over N_c^2} \right),
\label{400}
\end{eqnarray}
in good agreement with the experimental result.

The $\Delta^+ \to p \gamma$ also can be described using a multipole expansion. 
The only multipoles which contribute are the magnetic moment (M1) and the 
electric quadrupole (E2).  These multipoles are related to the helicity 
amplitudes 
by~\cite{PDG}
\begin{eqnarray}
A_{1/2} &=& -{1\over 2}\left( M1+3 E2 \right) , \nn
A_{3/2} &=& -{\sqrt 3 \over 2}\left( M1- E2 \right) . 
\label{500}
\end{eqnarray}
The total decay rate $\Gamma$ in Eq.~(\ref{100}) depends on
\begin{eqnarray}
\left[ \abs{A_{1/2}}^2 + \abs{A_{3/2}}^2 \right] &=& 
\left[ \abs{M1}^2 +3 \abs{E2}^2 \right].
\label{510}
\end{eqnarray}
The experimental value for the ratio $E2/M1$ is
\begin{eqnarray}
{E2 \over M1} &=& -0.025 \pm 0.005 \ .
\label{600}
\end{eqnarray}
We show that the large-$N_c$ prediction for $E2/M1$,
\begin{eqnarray}
{E2 \over M1} &=& \mathcal{O}\left( { 1 \over N_c^2} \right) \ , 
\label{700}
\end{eqnarray}
is equivalent to the prediction Eq.~(\ref{400}) for the ratio of helicity 
amplitudes.  The experimental measurement Eq.~(\ref{600}) is consistent 
with the prediction that $E2/M1$ is order $1/N_c^2$.

The large-$N_c$ expansion for baryons gives an expansion in powers of  $1/N_c$
for operators with definite spin and flavor transformation 
properties~\cite{dmj,djm,lm,cgo}.  In the $N_c \to \infty$ limit,
the baryons are infinitely  heavy and can be treated as static~\cite{jm}, with
recoil effects included as  $1/N_c$ corrections~\cite{prm}.  The static
kinematics of the large-$N_c$ limit makes it natural to use  large-$N_c$
methods  to evaluate the multipole amplitudes, rather than the helicity
amplitudes.

The $\Delta^+ \to p \gamma$ transition is an $I=3/2 \to I=1/2$ transition 
mediated by the isovector part of the electromagnetic current.  The isovector 
electromagnetic current operator can be expanded in powers of the  photon
momentum (the multipole expansion), and is proportional to 
\begin{eqnarray}
\left(J_{EM}\right)^{ia} &\propto& \mu^{ia} + Q^{(ij)a}k_\gamma^j + \ldots
\label{710}
\end{eqnarray}
where $i$ and $j=1,2,3$ are vector spin indices, and $a=1,2,3$  is a vector
isospin index. In Eq.~(\ref{710}), $\mu^{ia}$ is the magnetic moment operator,
which is an $(J,I)=(1,1)$ operator, and $Q^{(ij)a}$ is  the quadrupole moment
operator, which is an $(J,I)=(2,1)$ operator.  The quadrupole operator
$Q^{(ij)a}$ is a symmetric, traceless tensor in spin indices $i$ and $j$.  The
$M1$ and $E2$ amplitudes are related to the matrix elements of the magnetic
moment and quadrupole moment operators through
\begin{eqnarray}\label{m1e2def}
      M1 &=& e {k_\gamma}^{1/2}\ \langle N\, \uparrow |\mu_z|
   \Delta\, \uparrow \rangle\ , \nonumber \\
      E2 &=& {1 \over {12}} e{k_\gamma}^{3/2}\ \langle N\, 
      \uparrow | Q_{20}|\Delta\, \uparrow \rangle\ ,
\end{eqnarray}
where $\mu_z=\mu^{ia}$ for $i=3,a=3$, and $Q_{20}$ is the $\ell=2,m_\ell=0$
component of $Q^{(ij)a}$ for $a=3$.

As shown in the above equation,  the $M1$ amplitude for $\Delta^+ \rightarrow
p \gamma$ is proportional to the matrix element of the $(J,I)=(1,1)$ operator 
$\mu^{ia}$ between $\Delta^+$ and $p$. In the $N_c \rightarrow \infty$ limit,
the states $\Delta^+$ and $p$ are both contained in the lowest-lying baryon
spin-flavor multiplet consisting of baryons with $J=I$. The isovector magnetic
moment operator has a $1/N_c$ expansion in terms of operator products of the
baryon $SU(4)$ spin-flavor generators  $G^{ia}$, $I^a$ and $J^i$.  Each
independent operator polynomial arises at a definite order in the $1/N_c$
expansion which is determined by its explicit factor of $1/N_c$ and the
implicit $N_c$-dependence of its operator matrix elements. The $1/N_c$
expansion of the magnetic moment operator is given by
\begin{eqnarray}
\mu^{ia} &=& \mu_1 G^{ia} + \mu_2 {J^i I^a \over N_c} + \mu_3 {1 \over N_c^2}
\left\{ J^2, G^{ia} \right\} +\ldots ,
\label{800}
\end{eqnarray}
where the coefficients $\mu_1$, $\mu_2$, $\mu_3$, $\ldots$ have an expansion 
in $1/N_c$ beginning at order unity, and the ellipsis represents higher-body
operators.  The complete set of higher-body operators are of the form ${\cal
O}_{n+2} = \left\{ J^2, {\cal O}_n \right\}$ where ${\cal O}_1 = G^{ia}$ and
${\cal O}_2 = J^i I^a$. At leading order in $1/N_c$, the $1/N_c$ expansion of
the magnetic moment operator can be truncated to the single operator $\mu_1
G^{ia}$. This truncation of the $1/N_c$ expansion yields the spin-flavor
symmetry relation between the $\Delta^+ \to p \gamma$ transition moment  and
the isovector combination of the $p$ and $n$  magnetic moments~\cite{magmom},
\begin{eqnarray}
\mu_{\Delta^+ p}
&=& {1 \over {\sqrt 2}} \left( \mu_p - \mu_n \right).
\label{900}
\end{eqnarray}
The matrix elements of the spin-flavor generator $G^{ia}$ are of order $N_c$, 
so the transition magnetic moment matrix element  for $\Delta^+ \rightarrow p
\gamma$ is order $N_c$\footnote{The gyromagnetic ratio, the magnetic moment
in units of $1/(2M_N)$, is of order $N_c^2$ since the nucleon mass $M_N$ is of
order $N_c$.}. This power counting is in accordance with the $I=J$
rule~\cite{ijrule,ij1nrule}. Note that the $M1$ amplitude is only order
$\sqrt{N_c}$ when ignoring an $N_c$-counting for the electromagnetic coupling
$e$, since the kinematic factor $\sqrt{k_\gamma}$ in the definition
Eq.~(\ref{m1e2def}) gives a suppression factor of $\sqrt{1/N_c}$. The leading
$1/N_c$ corrections to Eq.~(\ref{900}) are produced by the  $J^i I^a/N_c$ and
$\left\{ J^2, G^{ia} \right\}/N_c^2$ operators which both yield corrections of
order $1/N_c^2$ relative to the leading order $N_c$ contribution since the
matrix elements of the baryon spin-flavor generators obey the $N_c$ power
counting $G \sim N_c$, $I \sim 1$ and $J \sim 1$.    

Since the $1/N_c$ expansion of the isovector magnetic moment is given by a
unique operator $G^{ia}$   to relative order $1/N_c^2$ and, as we will show,
the $E2$ amplitude is  order $1/N_c^2$ relative to the $M1$ amplitude, the
ratio of the helicity amplitudes  $A_{3/2}/A_{1/2}$ is predicted at leading
order in the $1/N_c$ expansion. The value of the prediction is given   in
Eq.~(\ref{400}). It is straightforward to show that all  higher order terms in
the $1/N_c$ expansion of the $M1$ amplitude  Eq.~(\ref{800}) do not affect
the  ratio $A_{3/2}/A_{1/2}$, although they do violate the $SU(4)$ symmetry
relation Eq.~(\ref{900}).

The $E2$ amplitude for $\Delta^+ \rightarrow p \gamma$ is proportional to the
matrix element of the $(J,I)= (2,1)$ operator $Q^{(ij)a} k_\gamma^j$ between
$\Delta^+$ and $p$. The $E2$ amplitude has one additional power of  the photon
momentum $k_\gamma$ relative to the $M1$ amplitude.  Since the mass difference
between $\Delta^+$ and $p$ is order $1/N_c$, $k_\gamma$ is order $1/N_c$. This
kinematic suppression  of $E2/M1$ by one power of $1/N_c$ is not sufficient to
explain the  smallness of the measured ratio, Eq.~(\ref{600}).  However, there
is an additional $1/N_c$ suppression beyond the kinematic suppression  due to
$k_\gamma$. The electric quadrupole amplitude is proportional to the matrix
element of  the $(J,I)=(2,1)$ operator $Q^{(ij)a}$ between $\Delta^+$ and
$p$.  The quadrupole moment operator has the $1/N_c$ expansion
\begin{eqnarray}
Q^{(ij)a} &=& c_1\left(q_u-q_d \right) {\left\{G^{ia}, J^j\right\}  \over N_c} + \ldots 
\label{1000}
\end{eqnarray}
where the operator $\left\{G^{ia}, J^j\right\}/N_c$ is understood to have its
trace subtracted, $c_1$ is an operator coefficient of order unity,
and $q_u$ and $q_d$ are the $u$ and $d$ quark charges.   The
electric quadrupole operator $Q^{(ij)a}$ has matrix elements of order unity at
leading order, since the matrix elements of the baryon spin-flavor generators
obey the $N_c$ power counting  $G \sim N_c$, $I \sim 1$ and $J \sim 1$. This
$1/N_c$ suppression is in accordance with the  suppression of $I\not=J$
operators  by ${\left(1/N_c\right)}^{|J-I|}$, as derived in~\cite{ij1nrule},
since $|J-I| = 1$ for the quadrupole operator. Thus $E2$ is order
$(1/N_c)^{3/2}$ due to the factor of  $(1/N_c)^{3/2}$ from $k_\gamma^{3/2}$ in
Eq.~(\ref{m1e2def}),  and $E2/M1$ is of order $1/N_c^2$ since $M1$ is order
$\sqrt{N_c}$.

Note that the isoscalar quadrupole operator has
the expansion,
\begin{equation}
       Q^{(ij)} = {c_0}\left(q_u+q_d \right) 
       {{\left\{J^i , J^j\right\}}\over N_c} + ... \ ,
\end{equation}
where $c_0$ is of order unity. Since the matrix element of $G^{ia}$ is of
order $N_c$ and of $J^i$ is of order unity, the ratio of the isoscalar to
isovector quadrupole moments is of order $1/N_c$ times $(q_u+q_d)/(q_u-q_d)$.
If one assumes that  $q_u=2/3$ and  $q_d=-1/3$, then $(q_u+q_d)/(q_u-q_d)=1/3$
and the ratio of the the isoscalar to isovector quadrupole moments is order
$1/N_c$. If, as in Ref.~\cite{bhl}, one chooses $q_u=(N_c+1)/(2N_c)$, and
$q_d=(1-N_c)/(2N_c)$,  then $(q_u+q_d)/(q_u-q_d)=1/N_c$ and the ratio of the
isoscalar to isovector quadrupole moments is order $1/N_c^2$, in agreement
with Ref.~\cite{bhl}.

One concrete realization of the large-$N_c$ spin-flavor symmetry of baryons in
QCD is given by the Skyrme model. Thus,  the Skyrme model  can be used as an
explicit check of the general, model-independent  large-$N_c$ results. For
instance, the Skyrme model predicts that the transition magnetic moment
$\mu_{\Delta N}=\langle\mu\rangle_{I=1}/\sqrt{2}$, where the isovector
magnetic moment is proportional to $f_\pi^{-1}e^{-3}$, and $e$ is the
coefficient of the Skyrme term. Since $f_\pi\sim \sqrt{N_c}$ and $e\sim 
1/\sqrt{N_c}$ , the transition magnetic moment scales like $N_c$~\cite{anw}.
Likewise, the Skyrme model predicts the transition quadrupole moment
$Q_{\Delta N} = (\sqrt{2}/10)\langle r^2\rangle_{I=1}$. Although the nucleon
isovector charge radius $\langle r^2\rangle_{I=1}$ diverges in the chiral
limit, it is independent of $N_c$ in the large-$N_c$ limit.  Therefore
$Q_{\Delta N}$ is of order $N_c^0$~\cite{an}. 

One can also apply the large-$N_c$ counting rules to the charge radii. For the
isovector charge radii, we have the expansion
\begin{equation}
      \langle r^2\rangle_{I=1}^a = a_1 (q_u-q_d){{I^a}} + ... \ , 
\end{equation} 
where $a_1$ is a coefficient of order unity. 
Thus $\langle r^2\rangle_{I=1}$  is of order $N_c^0$, consistent with the
Skyrme model prediction. For the isoscalar charge radii, 
\begin{equation}
      \langle r^2\rangle_{I=0} = a_0 (q_u+q_d) N_c  + ... \ , 
\end{equation} 
where $a_0$ is a coefficient of order unity. 
The ratio of the isoscalar to isovector charge radii is of order
$N_c (q_u+q_d)/(q_u-q_d)$, in agreement with Ref.~\cite{bl}. The above results
also are consistent with the picture that the baryon radius is of order $N_c^0$
in the large-$N_c$ limit. The charge radius $\langle r^2\rangle$ 
for a quark is of order unity. The
isoscalar charge radius is the sum of $\langle r^2\rangle$ over $N_c$ quarks,
and is of order $N_c$, whereas the isovector charge radius is the sum of
$\langle r^2\rangle$ over the $(N_c+1)/2$ $u$-quarks minus the 
sum of $\langle r^2\rangle$ over the $(N_c-1)/2$
$d$-quarks. The order $N_c$ pieces cancels in the difference, 
so the isovector charge radius is order one.

\acknowledgments

E.J and A.M. were supported in part by the Department of Energy under grant
DOE-FG03-97ER40546. X. J. thanks T. Cohen for useful discussions, and is 
supported in part by the DOE grant DOE-FG02-93ER40762.

\end{document}